\documentclass[%
rsi,%
superscriptaddress,
 amsmath,amssymb,
 reprint,%
]{revtex4-1}

\usepackage{graphicx}
\usepackage{dcolumn}
\usepackage{bm}
\usepackage{color}

\begin{document}

\preprint{AIP/123-QED}

\title[Mixmasters that work]{Quasi-isotropic cycles and non-singular bounces in a Mixmaster cosmology}

\author{Chandrima Ganguly}
\email{chandrima.ganguly@dartmouth.edu}
 \affiliation{Lindemann Fellow, Department of Physics and Astronomy,Wilder Laboratories, Dartmouth College, Hanover NH03755, USA}
\author{Marco Bruni}%
 \email{marco.bruni@port.ac.uk}
\affiliation{ 
Institute  of  Cosmology  and  Gravitation,  University  of  Portsmouth,
Dennis  Sciama  Building,  Burnaby  Road,  Portsmouth  PO1  3FX, UK
}%

\date{\today}

\begin{abstract}
A Bianchi IX Mixmaster spacetime is the most general spatially homogeneous solution of Einstein's equations and it can represent the space-averaged Universe. We introduce two novel mechanisms resulting in a Mixmaster Universe with non-singular bounces which are quasi-isotropic. Matter with a non-linear equation of state allows these bounces. Using a negative anisotropic stress successfully isotropises this Universe and mitigates the well known Mixmaster chaotic behaviour. The Universe can be an eternal Mixmaster, going through an infinite series of quasi-isotropic cycles separated by bounces.
\end{abstract}

\maketitle

\paragraph{\textbf{Introduction}}
Cosmological parameter extraction from Planck \cite{Planck1,Planck3,Planck4} and other studies \cite{saadeh} place tight constraints on the anisotropy of the current day Universe, $(\sigma _S/H)_o \sim 10^{-11}$, where $\sigma_S$ represent the contribution of the (scalar part of) shear anisotropic expansion, and $H$ the isotropic expansion rate. A main problem of classical cosmology is that a Big Bang singularity is quite ubiquitous \cite{hawking_singularity, israel, belinski_henneaux_2017}. Inflation through exponential expansion is able to propose a mechanism of diluting out any initial anisotropy, homogeneity and curvature. Most importantly, the inflationary scenario is predictive, as it provides the setting to generate the observable fluctuations in the cosmic microwave background and in the matter distribution. However there remain shortcomings within this paradigm. For example many models can't be extrapolated backwards, unable to escape a Big Bang. The energy scale of inflation can also be high enough for a complete theory of quantum gravity to be required \cite{energyScaleOfInflation}. All cosmologically interesting fluctuation modes may have originated from a trans-Planckian zone of ignorance - the physics of which would significantly impact the predicted spectrum of cosmological perturbations \cite{trans-Planckian}.
\paragraph*{}
The scenario proposed in this paper does away with this problem by producing an alternate `beginning' story to the observable Universe - a successful bounce which solves the issue of the Big Bang singularity and the trans-Planckian problem \cite{review2}, followed by a quasi-isotropic expansion. 
\paragraph*{}
We assume General Relativity (with units $c=8\pi G=1$) and a matter source with a non-linear equation of state (EoS)  \cite{bruni1,bruni2}
\begin{equation}\label{eq:non-linear equation of state}
    P= P_o + \alpha \rho + \frac{\epsilon}{\rho_c} \rho ^2\;,
\end{equation}
relating the pressure $P$ and the energy density $\rho$.
For the most general case of an anisotropic fluid conservation of energy is guaranteed by 
\begin{equation}
    \dot{\rho}=-3 H(\rho + P) - \pi _{\mu\nu}\sigma ^{\mu\nu},\label{eq:gencons}
\end{equation}
where $\pi_{\mu\nu}$ is the anisotropic stress.
Here $\rho_c$ is the characteristic energy scale at which  the effect of the non-linearities in the EoS becomes relevant. $P_o$ can play the role of an effective cosmological constant \cite{bruni1};  in this paper we assume $P_o =0$ and  $\epsilon = -1$. Then, for the perfect fluid case, this  gives an effective cosmological constant  
\begin{equation}\label{rhoL}
\rho _{\Lambda}=(\alpha +1)\rho_c\,,
\end{equation}
i.e.\ a stationary point of (\ref{eq:gencons}), so that it must be $\rho<\rho_\Lambda$, or otherwise one gets a ``phantom behaviour'', i.e.\ $\dot{\rho}>0$ during expansion ($H>0$) \cite{bruni1}. In other words, in our scenario we have a \textit{high-energy cosmological constant} for $\pi_{\mu\nu}=0$. Choosing $\alpha = 1/3$ to correspond to the radiation case at low energies,  when the quadratic term is negligible, we have therefore a maximum value $\rho_{\Lambda}=4\rho_c/3$ for  $\rho$.  A closed (positively curved space) Friedmann-Lema\^itre-Robertson-Walker (FLRW)  universe with this EoS  can undergo cycles with  non singular bounces, as shown  in \cite{bruni1}; here we want to use this EoS in the context of a Bianchi IX cosmology, generalising \cite{bruni1,bruni2}. In this scenario as long as $\rho _{\Lambda}$ is below the Planck energy scale the universe undergoes a bounce at energies where one can use classical General Relativity. Other components such as standard matter can be introduced, but here we restrict the analysis to the bare-bones of the scenario, to make it simpler.
\paragraph*{}
The next question addressed in this work is that of growing anisotropies in the contracting phase of a bouncing cosmology. We introduce a novel isotropisation mechanism using a negative anisotropic stress. This has been studied in expanding universes in \cite{misner_aniso}. A negative $\pi_{\mu\nu}$ successfully mitigates the growth of expansion anisotropy, as well as the chaotic behaviour that is ubiquitous in spatially homogeneous spacetimes with anisotropic $3$-curvature. This is an improvement from the use of the unphysical linear EoS $p \gg \rho$ that has been used in the literature so far \cite{Lidsey2005,burd}.
\paragraph*{}

\paragraph{\textbf{The Bianchi IX universe}}
To understand the behaviour of the most general kind of anisotropies - both in expansion as well as in the $3$ curvature - we turn to the Bianchi IX models. These are the most general anisotropic, spatially homogeneous spacetimes \cite{LLif}. When one considers solutions of Einstein's equations of this type,  under standard assumption on the matter content \cite{belinski_henneaux_2017}, typically these models approach a singularity where matter becomes negligible, with the anisotropic $3$-curvature driving the scale factors to undergo infinite chaotic oscillations over a finite time interval \cite{bkl,LLif,Deruelle,Sopuerta}. This chaotic behaviour is actually an attractor, called the Mixmaster attractor \cite{mis}.  An important reason why this class of cosmological models is interesting is because they contain the closed FLRW model as their isotropic sub-case. Universe models having positive spatial curvature have always been of particular interest as cycling solutions have been found in the isotropic case sourced by a linear EoS fluid \cite{bruni1,Btsagas} and by our present quadratic EoS \cite{bruni1}. In the Bianchi IX models the sign of the $3$-curvature is crucial: only when they are close enough to isotropy the $3$-curvature is positive and  re-expansion can occur. Sufficient isotropisation must therefore take place before the Universe can re-expand after collapse. Attempts at isotropising the Bianchi IX cosmology and mitigating the chaotic behaviour have mostly been focused on introducing stiff ($P=\rho$) \cite{Belinskii1972} or super-stiff ($P\gg \rho$) \cite{Lidsey2005,burd} matter. Adding a super-stiff fluid does not always seem to work as the existence of super-stiff anisotropic stress causes a faster growth of the energy density in the shear anisotropies \cite{me1}. We study the effects of the non-linear EoS \eqref{eq:non-linear equation of state} and the negative anisotropic stress by numerically integrating the Einstein's field equations and the conservation equation for the energy momentum tensor $T^{\mu}_{\nu}$ for  these cosmological models.

\paragraph*{}
In general, the metric for a homogeneous spacetime is given by 
\begin{equation}\label{eq:metric}
    ds^2 = -dt^2 + \gamma_{ab}\omega^a\omega ^b.
\end{equation}
Here $\omega^a$ are the one-forms for the triad basis in which the spacetime is defined. They are general functions of the spatial coordinates. The $\gamma _{ab}$ are the metric components in this triad and are functions of time only, as the spacetime is homogeneous. For our purposes, the metric in the triad frame is explicitly given by $\gamma_{ab}=\mathrm{Diagonal Matrix}[a(t)^2, b(t) ^2, c(t) ^2]$ where $a(t)$, $b(t)$ and $c(t)$ will be taken to be the dimensionless scale factors of the universe in the three spatial directions. The isotropic FLRW subcase is given by $a(t) = b(t) =c(t)$.
\paragraph*{}
We can write the equations of motion for this system in the triad basis in terms of $\gamma _{ab}$ and its time-derivative, the extrinsic curvature $\kappa_{ab}=\frac{\partial}{\partial t}\gamma_{ab}$\cite{LLif}. Thus, mathematically the problem is reduced to the study of a non-linear dynamical system of ordinary differential equations.

The Bianchi IX universe with matter has been found to undergo rotation of the triad frame axes themselves. The triad frame itself is part of the dynamics and the rotation can be parametrised in terms of the Euler angles $\theta$, $\phi$ and $\psi$. For our present purposes however, we are not interested in matter that exhibits non-comoving velocities or vorticities (see \cite{me3} and Refs.\ therein). Thus the stress-energy tensor $T_{\nu}^{\mu}$ is diagonal. Einstein equations then imply  that the non-diagonal components of the Ricci tensor are zero. It has been shown \cite{belinski_henneaux_2017} that this also implies that the Euler angles $\theta$, $\phi$ and $\psi$ are constant and frame rotation can be disregarded in this case.

We also introduce the following variable 
\begin{equation}\label{xyH}
    3x = \frac{\dot{a}}{a}-\frac{\dot{b}}{b},\;\;
    3y = \frac{\dot{a}}{a}- \frac{\dot{c}}{c},\;\;
    3H = \frac{\dot{a}}{a} + \frac{\dot{b}}{b} + \frac{\dot{c}}{c}.
\end{equation}
These variable definitions are useful: $3H$ gives the overall expansion of the volume of the universe and $x$ and $y$ are directly related to the shear $\sigma$ (made of the logarithmic derivatives of $a$, $b$ and $c$). They allow us to track the growth of anisotropy.  We choose a fluid 4-velocity $u^{\mu}= (1,0,0,0)$ and consider a diagonal stress energy tensor given by $T^{\mu} _{\nu} = \mathrm{Diagonal Matrix}[-\rho, P, P, P]$, where $P(\rho)$ is given by \eqref{eq:non-linear equation of state}. Therefore, using   Eq.\ \eqref{eq:non-linear equation of state} in  \eqref{eq:gencons}, together with Eqs.\ \eqref{xyH} and three equations for $x$, $y$ and $H$, 
 we have a dynamical system consisting of seven coupled first-order ordinary differential equations for seven variables.  

\paragraph{\textbf{Characteristic scale of the problem}}

The non-linear EoS \eqref{eq:non-linear equation of state} is characterised by the energy scale $\rho_c$. We can then introduce new dimensionless variables

\begin{equation}
 \tilde{x}=\frac{x}{\sqrt{\rho_c}},\;\;\;\tilde{y}=\frac{y}{\sqrt{\rho_c}},\;\;\;\tilde{H}=\frac{H}{\sqrt{\rho_c}},\;\;\; \tilde{\rho}= \frac{\rho}{{\rho _c}}\;,
 \end{equation}
 which amounts to introduce a dimensionless time $\eta=\sqrt{\rho_c}\, t$ \cite{bruni1}. Then,
 rewriting time derivatives in the equations of motion in terms of $\eta$
 the dynamics can made completely dimensionless and independent from $\rho_c$. Equivalently, for the purposes of our computations, we can use  $\rho_c$ as our unit,  setting $\rho_c = 1$ in our equations. However $\rho_c$ is related to the initial conditions and needs to be re-introduced in order to get physical results. We will comment on  the effect of this in a later section.

\paragraph{\textbf{Results from introducing only the non-linear fluid}}

\begin{figure}
    \centering
     \includegraphics[scale=0.5]{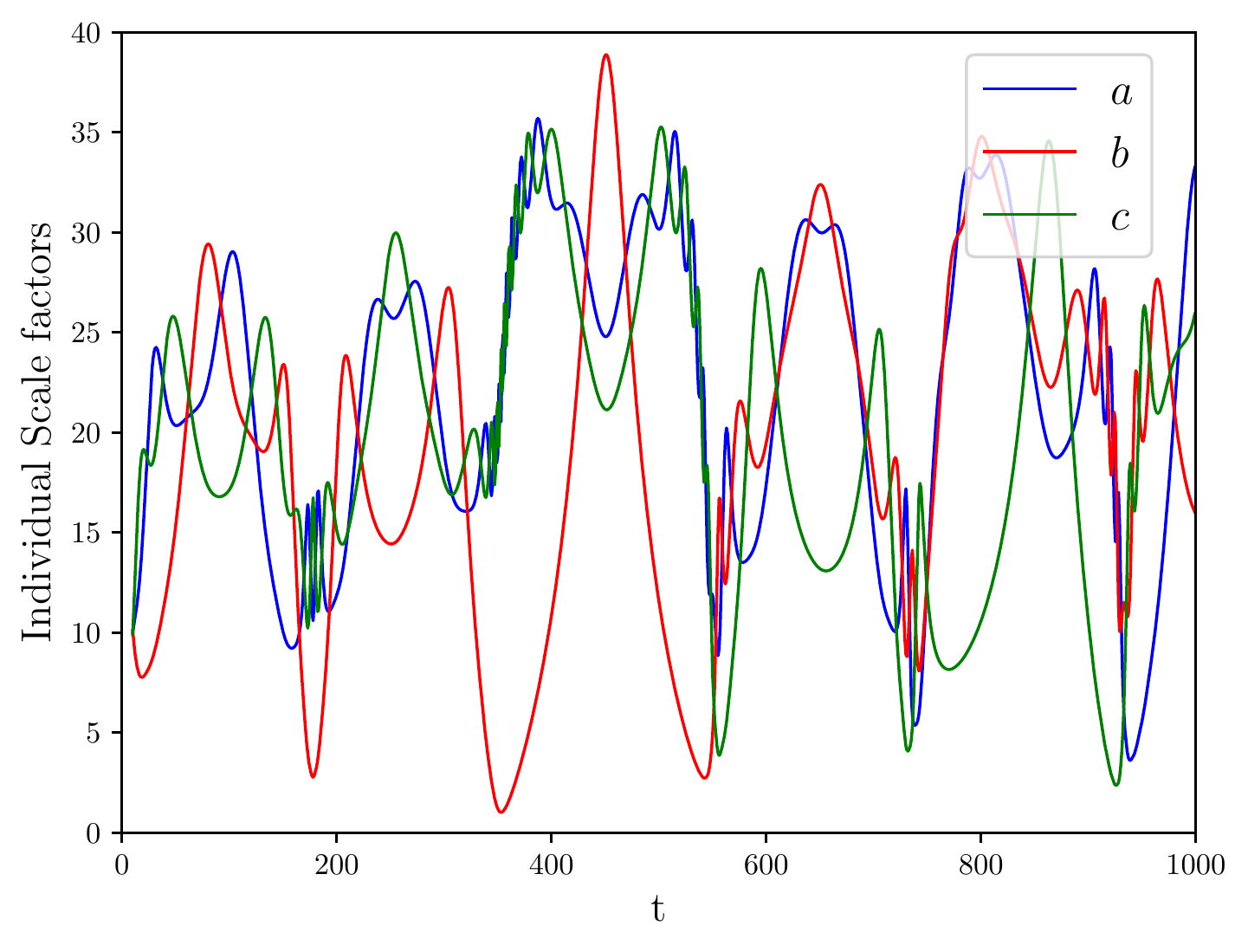}
     \caption{Blue, orange and green lines denote the scale factors $a$, $b$ and $c$, normalised to the curvature scale. The model is sourced by the non-linear equation of state $\frac{1}{3}\rho-\rho ^2$, with no anisotropic stress. Time is expressed in units of $\rho _c$.}
     \label{fig:scalefactorsNoAniso}
   \end{figure}

\begin{figure}
        \centering
         \includegraphics[scale=0.5]{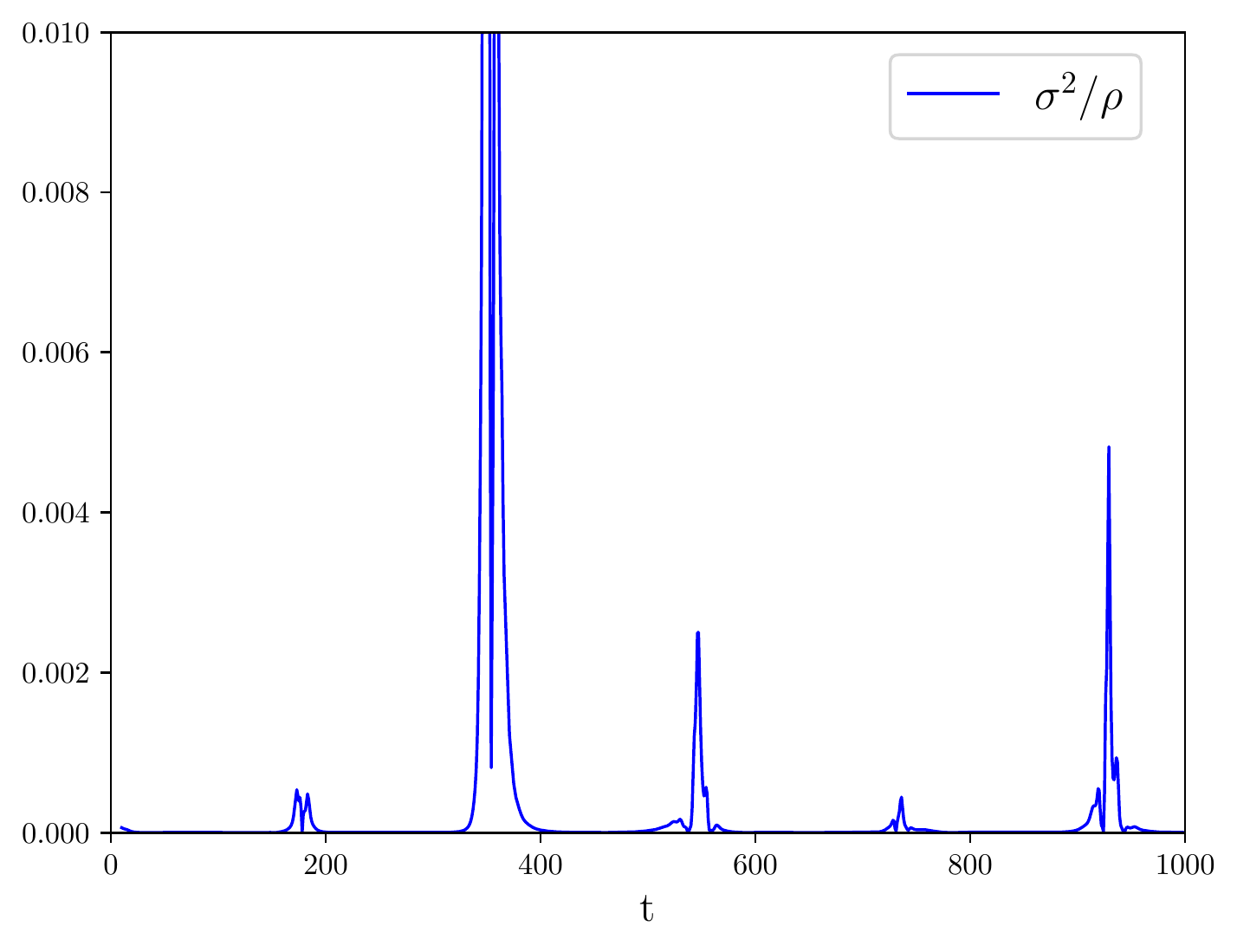}
    \caption{Plot of  the normalised dimensionless shear $\sigma ^2/\rho$. The model is sourced by the non-linear equation of state $\frac{1}{3}\rho-\rho ^2$, with no anisotropic stress.}
     \label{fig:shearNoAniso}
 \end{figure}

The results of the numerical integration in Fig.\ \ref{fig:scalefactorsNoAniso} reveal that the individual scale factors undergo several oscillations as expected and the volume scale factor undergoes bounces of similar height. This behaviour is reminiscent of the analysis done by \cite{zardecki} for linear equations of state, while the negative quadratic term in Eq.\ \eqref{eq:non-linear equation of state} is similar to the effective term in Loop Quantum Cosmology \cite{lqc_review}. 
The shear and the $3$-curvature shoot up at the minima but remain small when the universe is at its maximum size. The energy density in the anisotropies $\sigma ^2$ is diluted by expansion and is the smallest at the expansion maxima and show peaks near the minima. However we see from  Fig.\ \ref{fig:shearNoAniso} that the evolution of the shear energy density has chaotic peaks. This implies that although the non-linear EoS \eqref{eq:non-linear equation of state} is successful in creating a non singular bounce, it is not as successful in isotropising the universe, in contrast with the flat Bianchi I case studied in \cite{bruni_isotropy}. This bouncing universe doesn't seem capable of satisfying the stringent observational constraints on isotropy.

\paragraph{\textbf{Introduction of negative anisotropic stress as a mechanism of isotropisation}}

 The anisotropy energy density given by $\sigma ^2 \sim \sigma _{ab} \sigma ^{ab}$ grows as $(\mathrm{Volume})^{-2}$. The solution proposed by ekpyrosis \cite{Khoury2001,Lehners2008,Steinhardt2002a} is to introduce a scalar field rolling down a steep negative exponential potential. The idea is that this field - also known as the ekpyrotic field- will have an effective EoS that is super-stiff i.e.\ $P\gg \rho$. 
 This effective super-stiff fluid will evolve much faster than $(\mathrm{Volume})^{-2}$ in a universe with contracting volume and so will be able to dominate over the anisotropy, and inhomogeneity in the contracting phase of a bouncing universe. When anisotropic stress are included which are themselves super-stiff they act as a source for growing shear and hence any ekpyrotic field will always have to compete with these growing anisotropies. This has been studied in detail in \cite{me1}. Furthermore, a universe that is highly anisotropic and has non zero spatial curvature - like the Bianchi IX universe - will not re-expand through a bounce after contraction as the anisotropies will dominate over the fields that drive the bounce \cite{review}. In the specific case of the Bianchi IX universe this will signal the onset of chaotic behaviour. By the same reasoning, a negative anisotropic stress should cause the shear to decrease. This has been seen in the context of shear viscosity in expanding anisotropic universes in \cite{parnovskii}.
 \paragraph*{}
 We investigate whether similar forms of negative anisotropic stress can be used as a novel isotropisation mechanism in the presence of both expansion as well as curvature anisotropies.
 We choose the form of the anisotropic stress to be 
\begin{equation}
\pi_{ab} = \kappa \rho ^{1/2} \sigma _{ab},
\end{equation}
$\kappa$ is a dimensionless constant that we choose to be negative $\kappa <0$. This form of the anisotropic stress is also useful as the effect of these stresses only become significant at sufficiently high energies near the bounce. The negative proportionality constant should lead to the reduction of the shear without having to take recourse to introducing a super-stiff fluid. The stress energy tensor now becomes 
$T^{\mu}_{\nu} =\mathrm{Diagonal Matrix}[-\rho, p+\pi_{11}, p+\pi_{22}, p+\pi_{33}]$.

\begin{figure}
    \centering
     \includegraphics[scale=0.5]{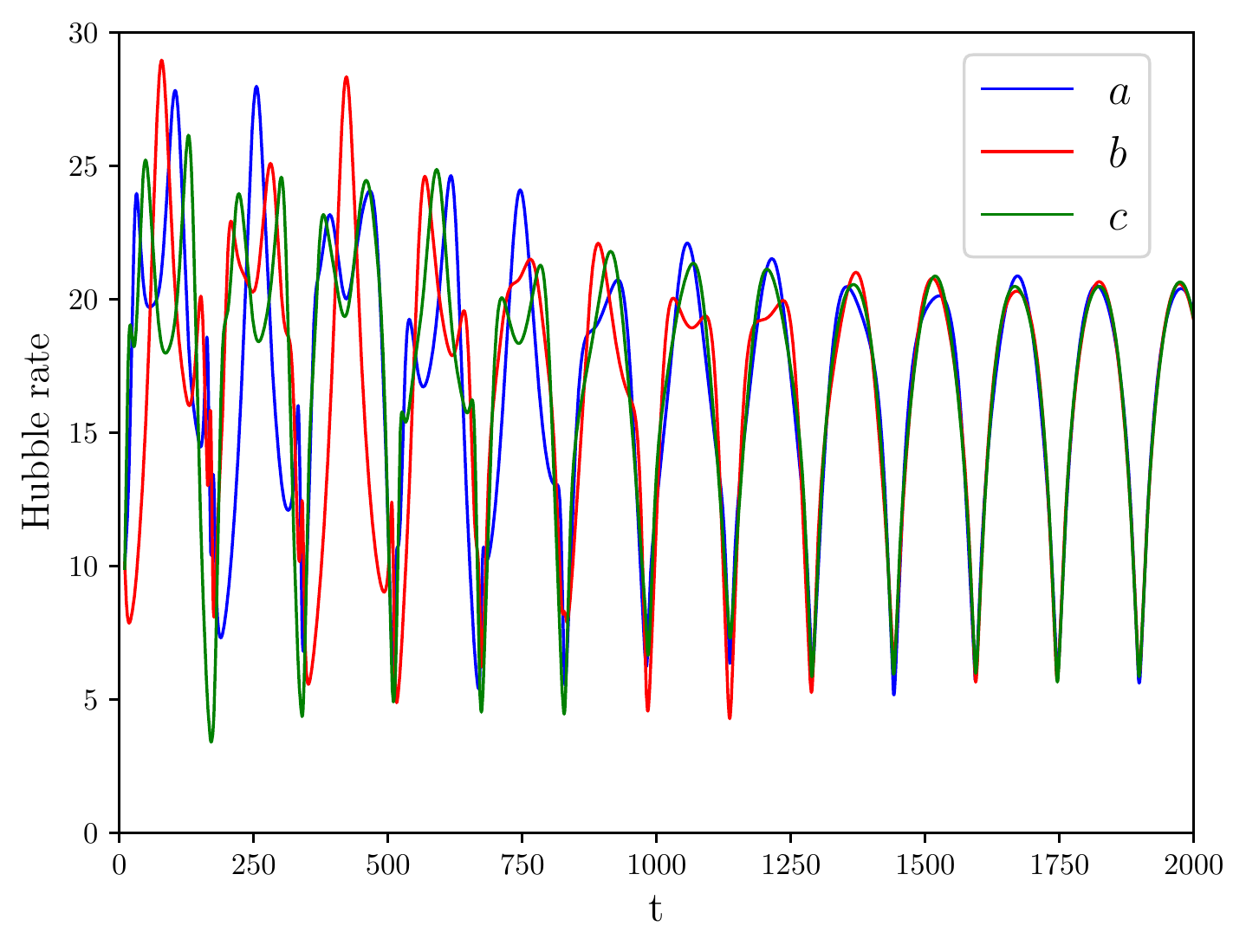}
    \caption{
    Blue, orange and green lines denote the scale factors $a$, $b$ and $c$, normalised to the curvature scale.
    Anisotropic stress $\pi _{ab} = \kappa \sqrt{\rho}\, \sigma_{ab}$ has been added to the fluid,  in addition to the non-linear equation of state for the isotropic pressure that was sourcing the model before. $\kappa = -4$ in this computation. Time is expressed in units of $\rho _c$.} \label{fig:scale_factors_with_stress}
    \end{figure}

\begin{figure}
    \centering
    \includegraphics[scale=0.5]{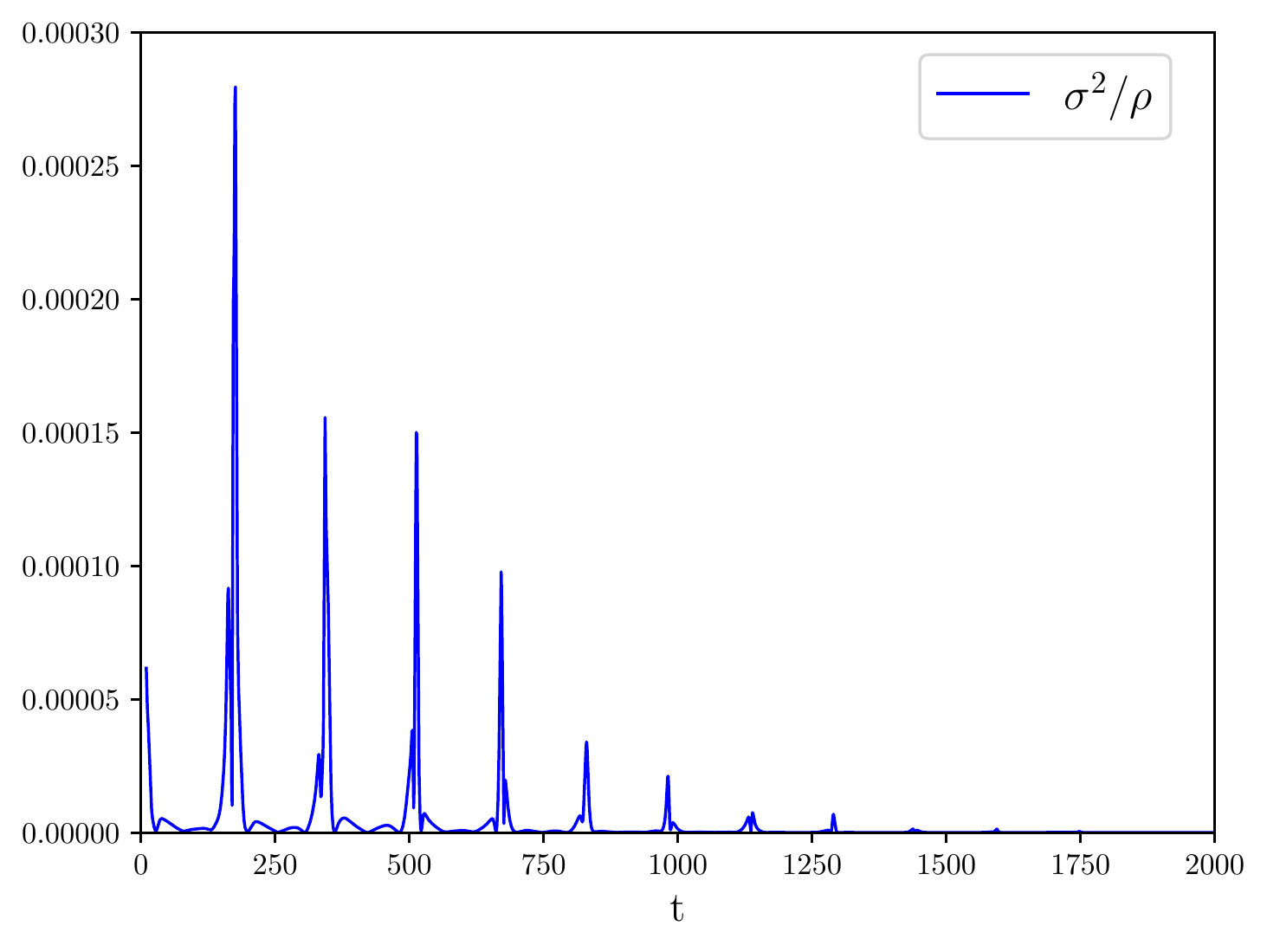}
    \caption{Normalised dimensionless shear $\sigma ^2/\rho$ for a model sourced by quadratic equation of state $P=\frac{1}{3}\rho - \rho ^2$ with the inclusion of anisotropic stress  $\pi _{ab} = \kappa \rho^{1/2} \sigma_{ab}$.  $\kappa = -4$ in this computation.}\label{fig:shear_with_anisoStress}
\end{figure}

We now solve the Einstein field equations in the Bianchi IX universe with the inclusion of this anisotropic stress. We find that the bouncing behaviour is still sustained with the bounces becoming more isotropic, see  Fig.\ \ref{fig:scale_factors_with_stress}. Furthermore we find, as shown in  Fig.\ \ref{fig:shear_with_anisoStress}, that the  peaks in the shear at the expansion minima of the model now decrease in height over time.
This seems to point to isotropy being achieved. It now remains to be seen if the chaotic behaviour that is an attractor in the Bianchi IX universe is also suppressed.

\paragraph{\textbf{Effect on Mixmaster chaos}}

The onset of chaotic behaviour in the Mixmaster system has been studied analytically, numerically and with invariant methods, e.g.\ see \cite{BChernoff},  \cite{zardecki, burd} and \cite{Cherubini:2004au} and references therein. By chaotic behaviour we mean that for a small change in initial conditions, the solution trajectories diverge exponentially from one another  during the course of the evolution, eventually  filling the phase space. The Principal Lyapunov exponent measures the rate at which the solution trajectories begin to diverge. A positive Principal Lyapunov index would signify the onset of chaos. Vice versa, a negative Principal Lyapunov index  signals the suppression of chaos.
We choose initial conditions containing some initial expansion as well as $3$-curvature anisotropy, but with this being small enough to resemble the isotropic Friedmann universe very closely. We then evaluate the  time evolution of the Principal Lyapunov index numerically to find that the Mixmaster chaos is completely mitigated, as the Principal Lyapunov index is negative and  of the order of $-2$. It is interesting to note that this happens even without the inclusion of a stiff fluid as observed in \cite{Belinskii1972} or a super-stiff field as shown in \cite{burd}. This is caused by the reduction in shear by the inclusion of the negative anisotropic stress as well as by the non-linear term in eq.\ \eqref{eq:non-linear equation of state}:  at energies close enough to $\rho_c$ this term  dominates and acts like an effective super-stiff fluid.

\paragraph{\textbf{Energy scale $\rho_c$ and size of the Universe at the bounce}}



In our scenario for the evolution of the Universe the non-linearity of the EoS \eqref{eq:non-linear equation of state}  depends critically from $\rho_c$. The effect of increasing its value is simply to increase the time period of the oscillations, and hence the amplification factor of the volume. 
Since our Bianchi IX model evolves toward the corresponding FLRW closed model \cite{bruni1}, we can use this to estimate the size of the Universe   at the bounce. Working now with a dimensionless scale factor $a(t)$ normalised to $a_o=1$ today and restoring units,  from \cite{bruni1} its minimum value $a_m$ at the bounce  is at least
\begin{equation}
a_m\geq \left(\frac{3 |\Omega_K|H_o^2                      c^2}{8\pi G (1+\alpha)\rho_c}  \right)^{1/2}
\end{equation}
for  $\rho$ bounded by $\rho_\Lambda$ (defined in Eq.\ \eqref{rhoL}), where $K=H_o^2|\Omega_K|$ is the curvature of the Universe, and $H_o$ and $\Omega_K$ are the today's measured Hubble parameter and curvature density parameters, respectively. The ratio of the size of the Universe today and its minimum is simply $a_o/a_m=a_m^{-1}$ and,  
if in turn we assume that $\rho_\Lambda$ is bounded by the Planck energy density, this ratio is of the order of 
\begin{equation}
\frac{a_o}{a_m}\lesssim \left(\frac{8\pi\, c^{5}}{3|\Omega_K|\hbar\, G H_o^2}\, \right)^{1/2}\;.
\end{equation}

This estimate would change with the inclusion of other matter components.


\paragraph{\textbf{Conclusion}} In this work, we have studied  Bianchi IX universe models sourced by a non-linear equation of state fluid. We find that with a negative quadratic term  in the  EoS \eqref{eq:non-linear equation of state} we can produce a series of non-singular bounces. These bounces and the subsequent cycles are reminiscent of the Mixmaster oscillations which are known to be chaotic in cases when the universe is sourced by fields obeying the strong energy condition (SEC), see  \cite{EllisChaotic} for an example. In our scenario, the field obeys the SEC at low energies but at energies close enough to $\rho_{\Lambda}$ the non-linear term that drive the bounce become important and the SEC can be violated. Crucially, \textit{the null energy condition is not violated} and $\dot{\rho}<0$ during expansion, avoiding a phantom behavior \cite{bruni1}. If the energy scale $\rho_\Lambda$ is sufficiently smaller than the Planck scale the evolution is purely classical and no quantum gravity corrections are needed.

We go a step further to propose a novel mechanism of isotropising the anisotropic closed Bianchi IX universe models by introducing a negative anisotropic stress. The effect is isotropisation as well as the mitigation or even suppression of the well-known Mixmaster chaos. This mechanism does not take recourse to the use of a super-stiff fluid reminiscent of the ekpyrosis scenario \cite{Khoury2001,Steinhardt2002}. 
\paragraph*{}
The net result of introducing the non-linear EoS \eqref{eq:non-linear equation of state} together with a negative anisotropic stress \eqref{eq:gencons} is that we obtain Bianchi IX cycling universe models that go through bounces followed by expansion and re-contraction, while remaining close to a FLRW  model with positive spatial curvature during expansion epochs. Thus these models, the most general homogeneous solutions of Einstein equations,  should satisfy  observational constrains \cite{Planck1,Planck3,Planck4}, specifically those on anisotropy  \cite{saadeh}, once all relevant matter components are included.

\paragraph*{}
For simplicity, here we have used a single fluid representing radiation at low energies, to demonstrate the mechanism of the bounce and isotropisation. Our results generalise easily to the inclusion of more matter components, in particular  a standard pressureless matter (giving a matter-dominate era) and dark energy, which are both subdominant to radiation and negligible once the evolution is close to a bounce. In future work, we will include such a standard matter component and we will investigate the effects of various dark energy scenarios on the type of cosmological models considered here.

\paragraph*{\textbf{Acknowledgements}} CG would like to acknowledge Marcelo Gleiser and John D. Barrow for many useful discussions. She would like to thank Dartmouth College Department of Physics and Astronomy for hosting her. She would also like to thank the English Speaking Union's Lindemann Trust for supporting this work. MB is supported by UK STFC Grant No. ST/N$000668$/$1$.


\bibliographystyle{apalike}
\bibliography{references-2}

\end{document}